\documentstyle[preprint,aps,epsfig,eqsecnum]{revtex}

\newcommand{\eqnapp}{\setcounter{equation}{0}%
      \renewcommand{\theequation}{A.\arabic{equation}}}%

\begin{document}

{\tighten

\preprint{\vbox{\hbox{FERMILAB--PUB--96/421--T}
                \hbox{HUTP-96/A055}}}

\title{\bf Large Charmless Yield in $B$ Decays and Inclusive $B$
Decay Puzzles}%

\author{Isard Dunietz, Joseph Incandela}
\address{\it Fermi National Accelerator Laboratory, P.O. Box 500,
Batavia, IL
60510}
\author{Frederick D. Snider}
\address{\it The Johns Hopkins University, Baltimore, MD 21218}
\author{Hitoshi Yamamoto}
\address{Harvard University, Cambridge, MA 02138}

\date{\today}
\maketitle
\begin{abstract}
In recent studies of inclusive $B$ decays, it has been suggested that
either $B$ mesons decay
much more copiously to final states with no
open charm than currently assumed, or $B(D^0 \to K^-\pi^+)$ has to
be reduced significantly.
This note takes the experimental $B(D^0
\to K^-\pi^+)$ at its face value and estimates $B(b\to$ no open
charm)
using complementary methods: one accounts for the
$c$ quark in $b\to c$ transitions, the other accounts for the
$\overline c$ quark in $b\to c\overline c s$ transitions.
Through cancellation of errors, the average gives our best estimate
of
$B(b\to$ no open charm), and the difference measures the
consistency. The results of the
methods are consistent with each other, strongly suggesting a
much enhanced $B(b\to$ no open
charm). This observation indicates that non-perturbative QCD effects are 
probably causing a sizable fraction of the $b\to c\overline cs$ 
transitions
to be seen as charmless $b\to s$ processes, contrary to smaller 
traditional expectations.
This mechanism has generally been overlooked and may explain the existing 
experimental data within the framework
of the standard model.
We then briefly discuss implications on
baryon production governed by $b\to c\overline cs$ processes,
rare hadronic $B$ decays and CP violation studies.

\end{abstract}

}

\newpage

\section{Introduction}
The puzzle of inclusive nonleptonic $B$ decays started out 
several years ago as the
discrepancy between the theoretical prediction and the experimental
measurement of the semileptonic branching 
ratio~\cite{altarellip,palmerstech,baffle,fwd}.  Theoretical analyses 
found it difficult to accomodate
$B_{s\ell}$ below 0.125~\cite{baffle},
while the experimental
value is~\cite{cleoslincl}
\begin{equation}\label{brsl}
  B_{s\ell} \equiv B(\overline B\rightarrow Xe^- \bar\nu ) = 0.1049\pm 
0.0046 \;,
\end{equation}
where $\overline B$ represents the weighted average of $B^-$ and 
$\overline B^0$.\footnote{The model-independent extraction of $B(b \to
\ell^- X)$ at $Z^0$ factories overlooked potentially significant 
effects~\cite{recal}.}
It was realized that there is a large uncertainty in the theoretical 
estimate of the $b\to c\bar cs$ rate.  The rate could increase due
to either a small charm quark mass or a failure of local
duality~\cite{baffle,fwd}, lowering the prediction for $B_{s\ell}$ down 
to the
experimental value. It would do so, however, at the expense of boosting 
the
charm multiplicity per $B$ decay ($n_c$) to around 1.3 which is 
significantly larger than the current
experimental value~\cite{CLEODX}:
\begin{equation}\label{nc}
   n_c = 1.10 \pm 0.05 \;.
\end{equation}
The puzzle was thus rephrased as the inability of theory and experiment 
to agree simultaneously on $B_{s\ell}$ and $n_c$~\cite{fwd}.
Subsequently, the inclusion of finite charm masses in
next-to-leading-order (NLO) calculations was found to 
enhance the $b\to c\bar cs$ rate by about 
30\%~\cite{rud,ncbagan,baganerr,voloshin}. 

The NLO calculations, however, are not complete (because penguin effects 
have not yet been included to NLO) and also suffer from large 
uncertainties due to charm quark mass,  renormalization scale and 
$\alpha_s(M_Z)$.  More significantly, the calculation is based on the 
underlying questionable assumption of
local quark-hadron duality. While duality assumes an inclusive rate based 
on 3-body phase space, the $b\rightarrow c\bar cs$ transitions proceed 
sizably as quasi-two body modes, which may enhance the inclusive
$b\rightarrow c\bar cs$ rate considerably~\cite{fwd}.

It was then shown~\cite{bsbsbar,bdy} that the uncertainty in
$b\to c\bar cs$ can be circumvented by noting that $B_{s\ell}$ and $n_c$ 
satisfy a linear
relation with the theoretical input of\footnote{Throughout this note, we 
define $d' \equiv V_{ud} d + V_{us} s$ and
$s' \equiv V_{cs} s + V_{cd} d$.}~\cite{rud}
\begin{equation}
\label{ruddef}
  r_{ud} \equiv \frac{\Gamma (b\to c\overline u d')}
                     {\Gamma (b\to ce\overline\nu )} = 4.0 \pm 0.4 \;.
\label{eq:rud}
\end{equation}
The above estimate of $r_{ud}$ is based on a complete NLO calculation
with finite charm quark mass and non-perturbative corrections up to
${\cal O}(1/m_b^2$).  Under the assumption of local duality, the error is 
dominated by the scale dependence and not by the uncertainties in quark 
masses and in $\alpha_s(M_Z)$.  Combining the accurately
measured $B_{s\ell}$ with the predicted $r_{ud}$ and with conventional 
assumptions regarding charmless yields in inclusive $B$ decays, 
Ref.~\cite{bdy} deduced
\begin{eqnarray}
  n_c &=& 1.30 \pm 0.05, \\
  B(b\rightarrow c\bar cs') &=& 0.32\pm 0.05\;. 
\end{eqnarray}
By simple accounting of the then observed `wrong-charm' yields or by 
studying the Dalitz plot distribution of the $b\to c\bar cs$ transition, 
a significant `wrong-charm' 
$\overline D$ production was predicted,
\begin{equation}\label{5}
  B(\overline B\rightarrow \overline DX) \approx 0.2 \;,
\end{equation}
where $\overline D$ represents $D^-$ or $\overline D^0$.
Subsequently, a sizable wrong charm $\overline D$ yield in $\overline B$ 
decays has been
observed by both CLEO \cite{moriond} and ALEPH \cite{barate} at 
approximately half the level as predicted.
The observation of the wrong-charm $\overline D$'s
does not alleviate the charm deficit problem, since the input
to the experimental value of $n_c$ is the total inclusive yield of $D$ and
$\overline D$ combined.

Refs.~\cite{recal,distw} tried to solve the charm deficit problem and
related issues by reducing
$B(D^0 \to K^-\pi^+)$ sizably below the current world average.
However, a recent precise measurement by ALEPH \cite{dokpialeph},
$B(D^0 \to K^- \pi^+) = 0.0390 \pm 0.0009 \pm 0.0012$,
agrees with previous measurements.  This indicated that
inclusive $B$ decays may not be well understood 
\cite{newphysics,dismontreal}, and
caused us to carefully reassess every input into the puzzle.

In this note, we take full advantage of newly available measurements,
in particular the flavor-tagged yields of $D$, $D_s$ and $\Lambda_c$,
and systematically identify the source of
the charm deficit to be the final
states with neither open $c$ nor open $\bar c$.
This branching fraction is denoted by $B(b\to \hbox{no open charm})$,
and is experimentally 
well-defined. It is the branching fraction to final states with no weakly 
decaying charmed 
hadrons, i.e. those states for which there can be no separate decay 
vertex resulting from weakly decaying charm.
This report then gives a plausible mechanism within the framework of
the standard model.

The rest of the paper is structured as follows:
In section II, we estimate $B(b\to \hbox{no open  charm})$
in two ways: method A focuses on the $c$ quark in
$b\to c$ transitions, and
method B focuses on the
$\overline c$ quark in $W\to \overline c s'$ transitions.
While method A uses experimental data and involves minimal theoretical 
input, method B requires a theoretical estimate for $r_{ud}$. Section III 
averages over
methods A and B (referred to as method C), which reduces errors 
significantly. Method C gives our best estimate of
$B(b\to \hbox{no open  charm})$, while
the difference between methods A and B checks the self consistency of
the analysis.
We find that the experimental data are self consistent and that
$B(b\to \hbox{no open  charm})$
is significantly larger than traditional
estimates. We then put forward a hypothesis that a sizable
component of $c\overline c$ pairs are seen as light hadrons and
not as open charm~\cite{palmerstech} through non-perturbative QCD effects.
Section IV discusses the systematics
of the analysis, which includes correlations
among the experimental and theoretical inputs.
Conclusions and some implications can be found in the last section.

\section{Two Ways of estimating
      $B(\lowercase{b\to\hbox{no open  charm}})$}

This report distinguishes flavor-specific branching
fractions -- $B(\overline B\to TX)$ and $B(\overline B\to \overline
TX)$ -- from the flavor-blind yield per B decay
\begin{equation}
   Y_T \equiv B(\overline B\to TX) + B(\overline B\to \overline
TX)\;.
\end{equation}
The branching fractions quoted by
experiments are the average number of particle $T$
per $B$ decay (weighted over charged and neutral $B$ productions).
When the particle $T$ is a charmed hadron, however,
it is safe to assume that the average number of particle per
decay is the same as the branching fraction.

$\overline B$ meson decays can be classified as $b\to c l
\overline\nu
(l=e,\mu,\tau)$,
$c\overline u d'$, $c\overline c s'$, $u\bar cs'$, and no 
charm.\footnote{`No charm'
indicates that there is no $c$ nor $\bar c$ quark in the final state
at quark level and includes $b\to u l \overline\nu$,
$u\overline ud'$, and charmless $b\to s'$
transitions.} Then, 
accounting for the weakly decaying charmed
hadrons originating from the $c$ quark in the $b \to c$ transitions,
we obtain
\begin{eqnarray}
   B(b &\to& \hbox{no open  charm}) = 1 - B(b\to u\bar cs')
                                             \nonumber  \\
                      &&   - B(\overline B\to DX)
                           - B(\overline B\to D_s^+X)
                           - B(\overline B\to N_c X)
   \qquad\hbox{(method A)},
   \label{eq:noopenc-A}
\end{eqnarray}
where
\begin{equation}
   B(b \to \hbox{no open  charm}) \equiv
             B(b\to \hbox{no  charm}) +
             B(\overline B\to (c\overline c)X)
     \label{eq:Bnoopenc}
\end{equation}
with $(c\overline c)$ being charmonia not seen as 
$D\overline DX$, and $N_c$ denotes any of the weakly decaying charmed 
baryons (namely,
$\Lambda_c,\: \Xi_c $ or $\Omega_c$).
The branching fraction $B(b\to u\bar cs')$ is small, and estimated
to be
\[
   B(b\to u\bar cs') = \left|{V_{ub}\over V_{cb}}\right|^2
\;\eta\;r_{ud} \; B(b\to c e \overline \nu) = 0.0035 \pm 0.0018\;,
\]
where $\eta \approx 1.3$ accounts for the larger QCD corrections in
$W \to \bar c s'$ transitions~\cite{ncbagan,baganerr,voloshin} 
with respect to those in $W \to \bar u d'$~\cite{rud}.  
Aside from this tiny correction, Method A involves
essentially no theoretical input.

The experimental inputs used in (\ref{eq:noopenc-A}) are
given in Tables I - III. Table I shows the flavor-blind
number of each particle type per B decay ($Y_T$) and
Table III shows the flavor-specific content of each yield.
Together, they provide flavor-specific branching fractions
needed in (\ref{eq:noopenc-A}).
We have used consistent values for the key branching
fractions of charm decays. The updated values are
summarized in Table II.
The experimental
value of $B(D^0 \to K^- \pi^+)$ is taken to be the 
new world average after the Warsaw '96 Conference~\cite{richman}:
\begin{equation}
   B(D^0 \to K^-\pi^+)= 0.0388 \pm 0.0010 \;.
\end{equation}
Both $B(D^+ \to K^-\pi^+\pi^+)$ and
$B(D_s^+ \to\phi\pi^+)$ are measured
model-independently and are proportional to $B(D^0 \to K^-\pi^+)$. The 
measured ratios are given in Table II.

Using the values in Tables I -- III and the definition
\begin{equation}
r_D \equiv \frac{B(\overline B\rightarrow \overline DX)}{B(\overline
B\rightarrow
DX)}\; ,
\end{equation}
the flavor-specific 
`wrong-sign' $\overline D$ ($\overline D^0$ or $D^-$) yield is 
\begin{equation}
  B(\overline B\to \overline DX) = Y_D \times {r_D\over 1+r_D}
       = 0.085 \pm0.025\qquad\hbox{(CLEO)}\;.
  \label{eq:DbarCLEO}
\end{equation}
The same quantity can be inferred from the ALEPH measurement of
$\overline B\to D\overline DX$~\cite{barate} to be (see Appendix)
\begin{equation}
  B(\overline B\to \overline DX) =
       0.145\pm0.037\quad\hbox{(ALEPH)}\;.
  \label{eq:DbarALEPH}
\end{equation}
The CLEO and ALEPH results are
consistent with each other within two standard deviations.
The agreement is mildly encouraging since they
have been measured using completely different methods.
The `right-sign' $D$ yield as well as the 
flavor-specific yields of $D_s$ and
$\Lambda_c$ are obtained similarly to (\ref{eq:DbarCLEO}).
The flavor-specific $D^+_s$ production in
$\overline B$ decays has been measured to be small by
CLEO~\cite{ds} (see Table III). This conclusion has been
confirmed by ALEPH~\cite{barate}.

The most accurate measurements regarding charmed baryon production in $B$ 
decays involve $\Lambda_c$ baryons.  In contrast, $\Xi_c$ production in 
$B$ decays involves large experimental uncertainties, and the $\Omega_c$ 
yield has not yet been observed.  Instead of the uncertain and 
nonexistent measurements, Refs.~\cite{recal,distw} inferred the inclusive 
$N_c$ yields by correlating them to the more accurately measured 
$\Lambda_c$ yields (see Appendix).  It predicted the $\Xi_c$ production 
to be drastically reduced with regard to the measured central 
value~\cite{browder}.
The drastic reduction can be traced back to a large enhancement in the 
absolute BR scale of $\Xi_c$ decays, a conclusion supported by recent 
work of Voloshin~\cite{VoloXic}.

We now turn to the second way (method B) of estimating 
$B(b \to \hbox{no open charm})$
which is to account for the $\overline c$ quark in $b\to c \overline
c s',\;u \overline c s'$
transitions. Noting that, apart from charmonia,
the $\overline c$ quark hadronizes to $\overline D$, $D_s^-$,
or $\overline N_c$, we obtain
\begin{eqnarray}
   B(b &\to& \hbox{no open  charm}) \nonumber \\
         &=& R - B(\overline B\to\overline DX)
               - B(\overline B\to D_s^-X)
               - B(\overline B\to\overline N_c X)
   \qquad\hbox{(method B)}\;.
   \label{eq:noopenc-B}
\end{eqnarray}
Here $R$ is the `remainder' of $\overline B$ branching fractions after 
reliable components have been subtracted:
\begin{eqnarray}
\label{Rdef}
   R &\equiv& B(b\to \hbox{no charm}) + B(b\to c \overline cs') +
B(b\to u \overline cs')
                                \nonumber \\
          &=& 1 - B(b\to c (e,\mu,\tau) \overline\nu)
                - B(b\to c \overline ud')          \nonumber \\
          &=& 1 - B(b\to c e \overline\nu)\; (2 + r_\tau + r_{ud}) \; .
\end{eqnarray}
The normalized tau semileptonic rate
\begin{equation}
  \label{rtau}
  r_\tau \equiv\frac{\Gamma (b\to c\tau\overline\nu )}{\Gamma (b\to
  ce\overline\nu )} = 0.22 \pm 0.02\;
\end{equation}
is reliably estimated by theory~\cite{rtau}, and is consistent with 
present
measurements. Using this as well as Eqs.~(\ref{brsl}) and (\ref{eq:rud}), 
one finds $R= 0.35 \pm 0.05$.
This result changes only minimally to
\begin{equation}
   \label{Rdiff}
   R=0.36 \pm 0.05\; ,
\end{equation}
when Pauli interference and $W$ annihilation effects are taken 
conservatively into account~\cite{neuberts,bbd}.
Our prediction (\ref{Rdiff}) for $R$ combines the most
accurate information available from both theory and experiment.

Using the experimental values from Tables I - III, we obtain
for methods A and B,
\begin{eqnarray}
   B(b \to \hbox{no open  charm}) &=&
        0.15 \pm 0.05\hbox{ (A) },\quad
        0.17 \pm 0.06\hbox{ (B) }\quad\hbox{ (CLEO)}
       \label{eq:CLEO}                \\
   & &  0.21 \pm 0.06\hbox{ (A) },\quad
        0.11 \pm 0.07\hbox{ (B) }\;\;\hbox{ (ALEPH\&CLEO)} .
       \label{eq:ALEPH-CLEO}
\end{eqnarray}
In (\ref{eq:ALEPH-CLEO}), we have used
$B(\overline B\to \overline DX)$ given by (\ref{eq:DbarALEPH}) and
\[
  B(\overline B\to DX) = Y_D - B(\overline B\to \overline DX)
\]
with all other inputs (including $Y_D$) being identical to those of 
(\ref{eq:CLEO}).

Fig. 1 shows the estimates of
$B(b \to \hbox{no open  charm})$
as functions of $B(D^0\to K^-\pi^+)$ using CLEO data only.
The experimental value of $B(D^0\to K^-\pi^+)$
is well within the overlap of the two bands, which
represent methods A and B.  This indicates self consistency of
the inputs.

\section{Best Estimate for $B(\lowercase{b\to\hbox{no open  charm}})$
     and Interpretations}

The errors in methods A and B are
highly correlated. For example, when
the ratio of wrong-sign to right-sign $D$'s ($r_D$) fluctuates
upward, the value given by A will increase while that given by B will 
decrease.
The best estimate of $B(b \to \hbox{no open  charm})$ can be obtained
by averaging over methods A and B, where the errors
due to flavor-specific fractions (namely, $r_D$, $r_{\Lambda_c}$ and
$f_{D_s}$; see Table III) cancel:
\begin{eqnarray}
   B(b \to \hbox{no open  charm})
             &=&0.5\;(1 + R - B(b\to u \overline cs') - Y_D - Y_{D_s}
- Y_{N_c})
          \qquad\hbox{(method C)} \nonumber     \\
        &=& 0.16 \pm 0.04\; \qquad\hbox{(CLEO).}
   \label{eq:noopenc-C}
\end{eqnarray}
The correlations are properly taken into 
account in the error estimation. The value is much larger than
the traditional estimate of $ B(b \to \hbox{no open  charm})$.

Eq.~(\ref{eq:Bnoopenc}) defines $B(b \to \hbox{no open  charm})$, where 
\begin{equation}
   B(b \to \hbox{no charm}) = 
    B(b\to u ({\rm no}\;\bar c)) + B(b\to s')\;.
\end{equation}
Here $B(b\to s')$ includes $b\to s'(ng, q \overline q)$ processes and 
interference effects.

 The $b\to u$
transitions are not large ($\sim1\%$) because of the small value of
$|V_{ub}/V_{cb}|$, while the
$b\to s'$ transitions have been argued to be
small due to the small Wilson coefficients of penguin 
operators~\cite{simma}.
Traditional estimates yield \cite{bdy}
\begin{equation}
 \label{nocest}
   B(b \to \hbox{no charm}) = 0.026 \pm 0.010\quad
 \hbox{(traditional guess)}.
\end{equation}
Conventional charmonia $(c\overline c)$ production in $B$ decays has been 
estimated to be~\cite{bdy}
\begin{equation}
  \label{ccbarest}
   B(\overline B \to (c\overline c) X) = 0.026 \pm 0.004\quad
  \hbox{(traditional guess)}.
\end{equation}
It used experimental measurements for $J/\psi , \psi', \chi_{c1}$,
and $\chi_{c2}$ together with theoretical estimates of other hidden
charmonia not yet detected. The $B(B\to\eta_c X,\eta_c' X)$ predictions 
used published calculations for decay constants of
$\eta_c,\eta_c'$ and related their yields to that of
$J/\psi$ assuming color-suppressed factorization, which cannot be 
justified theoretically~\cite{buras}.
The total yield of other charmonia
including those not expected from factorization (such as 
$h_c$ and $\chi_{c0}$)
were assumed to be $1.2 B(B \to \chi_{c2} 
X)$.\footnote{Eq.~(\ref{ccbarest}) is clearly unreliable and
 one should search for not only $\eta_c$ in
$B$ decays~\cite{browder} but also for other $(c\bar c)$, such as
$\eta_c', \chi_{c0},
h_c , {}^1D_2 , {}^3D_2 $.}
Adding up (\ref{nocest}) and (\ref{ccbarest}), we obtain
\begin{equation}
  \label{noopencest}
  B(b\to \hbox{no open  charm}) = 0.052 \pm 0.011\; \quad
  \hbox{(traditional guess)}.
\end{equation}

The traditional estimate (\ref{noopencest}) falls far
below $0.16\pm0.04$. Though estimate (\ref{ccbarest}) is unreliable due to
the questionable assumptions made, we do not expect the true conventional 
$(c \overline c)$ production to be large
enough to explain the bulk of the discrepancy.  What could be the
source of such a
large enhancement of $B(b \to \hbox{no open  charm})$?

New physics is one possible solution~\cite{newphysics}.  But before 
drawing that conclusion, all standard model explanations, including 
non-perturbative effects, have to be ruled out.  We hypothesize that 
non-perturbative effects could cause a significant fraction of 
$c\overline c$ pairs produced in $B$ decays to be seen as light 
hadrons~\cite{palmerstech}. This hypothesis does not modify the
previous analysis since the
expressions for methods A and B 
[Eqs.~(\ref{eq:noopenc-A}) and (\ref{eq:noopenc-B})]
allow for $c\overline c$ transformations to light hadrons and only assume 
that singly produced charm decays weakly.

How realistic is such a scenario?
The QCD corrected operator responsible for the $b\to c\overline cs$
transition can be written as (neglecting the small conventional
penguin contributions)
\begin{equation}
  \label{q2}
  2 c_2 (\overline s T^a b)_{V-A} (\overline c T^a c)_{V-A}
    + \big( c_1 +\frac{c_2}{N_c} \big)
   (\overline sb)_{V-A} (\overline cc)_{V-A} \; .
\end{equation}
The estimate for the coefficient of the color-singlet term
$(c_1 + c_2/N_c)$ ranges from 0.10 to 0.25 and is much
smaller than $c_2 \approx 1.1$~\cite{buras}.
Thus, the $c\overline c$ quark pair is 
produced dominantly in a color-octet configuration. This means that
the $c\bar c$ pair can annihilate into a single gluon. Such
effects, however, have already been included in the 
short-distance, perturbative calculations of 
$b\to s'$. Whatever may enhance the
$c\bar c$ transformation into light hadrons should then be due to 
non-perturbative effects.

One possibility is that light hadrons have a non-negligible
$c\bar c$ component~\cite{sea,valence}.  The part of the light hadron 
[$\pi, \rho, K^{(*)}$, etc.] wavefunction  that involves intrinsic charm 
will have maximal amplitude at minimal off-shellness and minimal 
invariant mass~\cite{sea}.  Thus it maybe significant that the $c\bar c$ 
pairs produced in $b \to c \bar c s$ transitions favor low invariant 
masses (see Figure 2).

Another candidate is a sizable production of $c\bar c g$ hybrids (denoted 
as 
$H_c$)~\cite{kuti,isgurpaton,hybrid,Hc,bcs} where the $c\bar c$ pair
is expected to be predominantly in a color-octet state.\footnote{We are 
grateful
to J. Kuti for pointing this out to us.}
Such hybrid states may couple strongly to the
color-octet $c\bar c$ pair produced in $\overline B$ decays governed by 
$b\to c\bar cs'$ transitions. 
The masses of the lowest lying $c\overline c$-hybrid mesons are  
predicted to
be above open charm threshold~\cite{kuti,Hc,bcs}. Still, their widths are
expected to be narrow because of selection rules that suppress the $H_c 
\to
D^{(*)} \overline  D^{(*)}$ transitions~\cite{isgurpaton,selrul}. The $H_c
\to D \overline  D^{**},D^{**} \overline D$ processes are kinematically
forbidden, except  for the reduced production of the broad $D^{**}$ mesons
with low  invariant masses.  Thus, such hybrid mesons may be 
seen significantly as light hadrons. At present, there is no
firm proof that this mechanism can account for the observed
enhancement of $B(b\to \hbox{no open  charm})$.
Non-perturbative QCD effects, however, are rich and poorly known.  We 
thus consider it important to investigate further theoretically and
experimentally whether a significant portion of $c\overline c$ pairs 
produced in $\overline B$ decays
could be seen as light hadrons.

\section{Systematics and correlations among observables}

The self consistency of inputs can be checked by 
taking the difference of the two methods which should equal zero:
\begin{eqnarray}
    B(b \to \hbox{no open  charm})\hbox{ (A)}
 &-& B(b \to \hbox{no open  charm})\hbox{ (B)}  \nonumber \\
      &=& -0.02 \pm 0.08\qquad\hbox{(CLEO)}   \\
      & &  \;\;\;0.10 \pm 0.10\qquad\hbox{(ALEPH\&CLEO).}
\end{eqnarray}
The CLEO data are clearly self consistent, but the
ALEPH data also are not inconsistent. 

Equivalently, equating methods A and B, we obtain a relation among
input parameters $B(D^0\to K^-\pi^+)$, $r_D$, 
$Y_T$, $r_{ud}$, etc. among which $r_{ud}$ is the only significant 
theoretical input
(other theoretical parameters are either reliable or small). This relation
can be used to check the self consistency of the inputs, or to solve for
one of the parameters in terms of all else. Solving for 
$B(D^0\to K^-\pi^+)$, $r_D$, and $r_{ud}$, one obtains
(using CLEO data only)
\begin{eqnarray}
    \label{eq:solve}
    B(D^0\to K^-\pi^+) &=& 4.0\pm0.5\%\;,\\
    r_D &=& 0.12 \pm 0.05\;,  \\
    r_{ud} &=& 4.1 \pm 0.7\;.
\end{eqnarray}
Note that the above determination of $r_{ud}$ uses experimental
inputs only. The fact that these values are consistent with the
input values themselves indicates that the inputs are
self consistent. We will first discuss the systematics of each input, 
and then examine the correlations among them.

\subsection{$Y_D$ and $Y_{D_s}$}
Could CLEO have badly mismeasured the coefficient (0.876 $\pm$ 0.037)
of $Y_D$ and/or the coefficient (0.1177 $\pm$ 0.0093)
of $Y_{D_s}$ (i.e., apart from $B(D^0\to K^-\pi^+)$, see Table I)?
In order for the best estimate of $B(b\to$ no open charm)
(method C) to
come down to the 5\%\ level, $Y_D + Y_{D_s}$ needs to increase by about 
20\%. That appears unlikely
since then the charm multiplicity in
$B$ decays as measured by
CLEO should be significantly different from recent
ALEPH~\cite{ccaleph} and
OPAL~\cite{ccopal} measurements which are given in Table
IV. Such a comparison is justified since the combined yields of 
$D, D_s, \Lambda_c$ in $b$-hadron decays at 
$Z^0$ and $\Upsilon (4S)$ factories are expected to agree
within existing experimental errors.
Table IV shows the consistency of the measurements. Also,
method A is more sensitive to the change in $Y_D$ and $Y_{D_s}$ than 
method 
B, and increasing $Y_D$ and $Y_{D_s}$ by 20\%\ 
results in a 2-sigma discrepancy
between the two methods evaluated at the nominal value of
$B(D^0\to K^-\pi^+)$.

\subsection{Charmed baryon yield}
We decided not to use the experimental
$\Xi_c$ data, and adopted a model prediction which gave 
branching fractions smaller than
the experimental values.
Even if we were to double the total charmed baryon yield in $B$ meson
decays, however, the result $B(b\to$ no open charm$)=0.14\pm0.04$ 
via method C would still be
significantly larger than the traditional estimate.
Thus, our conclusion is not sensitive to the uncertainty
in the charmed baryon yield.

\subsection{Wrong-sign/right-sign ratio of $D$ meson ($r_D$)}
The wrong-sign/right-sign ratio of $D$ mesons, $r_D$, still
has large uncertainties. The method currently employed by CLEO uses 
angular
correlations between a high energy lepton and a $D$ meson to
separate the cases where the $D$-lepton pair comes from the same
$\overline B$ meson or different $\overline B$ mesons. At low $D$
momenta, however,
the angular correlation is smeared out and it is difficult to
distinguish the two cases.
The ALEPH measurement fully reconstructs both charmed mesons from a
single $B$ thus avoiding such systematics, but suffers from low
statistics.  $Z^0$ factories
should be able to determine $r_D$ more accurately by measuring the 
inclusive yield of \underline{single} $D$'s in $b$-enriched data samples 
that are optimally flavor-tagged.  Neither flavor-tagging nor $B^0 - 
\overline B^0$ mixing corrections would be necessary, if a large charged 
$B$ sample could be isolated.

\subsection{$r_{ud}$}
Another possibility is that theory is unable to predict
$r_{ud}$ reliably. Local quark-hadron duality may not hold.  Once local 
duality is assumed, the most important
uncertainty lies in the choice of scale $\mu$, as mentioned 
earlier~\cite{rud}.  Figure 3
demonstrates a troubling aspect of the calculation. Contrary to 
expectation,
there is no significant reduction in sensitivity on $\mu$ when going from
leading-order to next-to-leading order. Maybe $r_{ud}$ has a 
significantly larger
uncertainty than currently appreciated. It is
gratifying to note that the recent measurements of wrong charm yields
allow the experimental determination of $r_{ud}$, which agrees with 
theory.

If the theoretical estimate of $r_{ud}$ is not to be trusted,
one has to rely on method A which does not depend on $r_{ud}$.
Note that the averaging used in method C reduces the
sensitivity to the uncertainty in $r_{ud}$.

\subsection{Correlation between $r_D$ and $B(D^0\to K^-\pi^+)$}
Figure 4 shows the hypothetical case of
$r_D = 0.20 \pm 0.03$, which agrees with the central value of the
ALEPH measurement [Eq.(\ref{eq:DbarALEPH})].  The lines would cross at
$B(b\to$ no open charm) = 0.06 and $B(D^0\to K^-\pi^+)$
= 0.032.  It demonstrates that increasing the wrong charm yield
makes $B(b\to\;$ no open charm) more
consistent with the traditional estimate.  The charm deficit would
disappear due to the lower value of $B(D^0 \to K^- \pi^+)$.
Thus if $r_D$ is measured to be around 0.2 with good accuracy, then one 
suspect would be a mismeasured $B(D^0 \to K^- \pi^+)$.  A more plausible 
culprit, however, would be a smaller $r_{ud}$ than theoretically 
predicted, as discussed next.

\subsection{$B(b\to$ no open charm) and $r_{ud}$ vs $r_D$}
Fig. 5 shows $r_D$ dependences of $r_{ud}$ and $B(b\to$ no open charm) 
(method A)
both of which use experimental inputs only.
If $r_D$ were small and around 0.05, one sees that $B(b\to$ no open charm)
is $\sim0.11\pm0.05$ which is within 1 sigma of the traditional 
estimate, and $r_{ud}\sim4.9 \pm 0.6$. The value of $r_{ud}\sim 5$ 
corresponds to $\mu\sim m_b/3$.\footnote{Using the BLM scale-setting 
method \cite{blm}, it has been
estimated that such small scales could be appropriate~\cite{lowscale}.}
This set of parameters would be more or less consistent with the
standard model without invoking new physics nor enhanced $c\bar c$
transformation into light hadrons. If on the other hand we take $r_D = 
0.20\pm0.03$, one
obtains $r_{ud} = 2.9\pm 0.6$, which 
disfavors small renormalization scales.
These discussions clearly show 
the importance of accurate measurements of $r_D$.

\section{Summary and Discussion}

Newly available flavor-tagged data made it possible to apply 
complementary methods to estimate $B(b\to$ no open charm).  Comparisons 
of the methods allowed us
to study correlations and self consistency of inputs.
$B(b\to$ no open charm) has been found to be much larger
than generally accepted.
The observation may indicate that non-perturbative effects cause an 
appreciable fraction of produced $\overline cc$ pairs in $B$ decays to be 
seen as light hadrons.

A large $B(b\to$ no open charm) could well be
the final missing piece in the puzzle
of the small charm multiplicity in $B$ decays and small $B(\overline
B\to X\ell\overline\nu )$.
The proposed mechanism of annihilation of
$\overline cc$
pairs could explain the low observed ratio of\cite{baryonc}
$$B(\overline B\to \overline N_c X)/B(\overline B\to N_c X). $$
The numerator is governed essentially by $b\to c\overline
cs'$ transitions, where a sizable fraction of $c \overline c$ pairs may 
not be seen as open charm thereby reducing the charmed baryon yield.
In contrast, the denominator is dominated by $b\to c\overline ud'$
processes which would result in single open charm.
The mechanism of $c \bar c$ transformation is also
consistent with the observed
significant surplus of
$K^-$ in inclusive $\overline B$ decays beyond conventional sources
and the measured large $K$-flavor correlation with $B$-flavor at time of
decay~\cite{inclk,browder}.

One way to measure $B(b\to$ no open charm) could use
a vertex detector which searches in a
$b$-enriched sample for a $b$-decay vertex and vetoes on
additional vertices from open charm.
In addition, one could then search for a kaon attached to the vertex.

If our predictions are confirmed, then many studies of rare $B$
decays and CP violation will have to be re-evaluated. Through 
non-perturbative
effects, amplitudes governed by $b\to d\;(s)$
transitions could have enhanced contributions governed by the 
combination of CKM matrix elements $V_{cb}V^*_{cd} \;(V_{cb} V^*_{cs})$. 
This indicates that the rate of $\overline B\to K^- \pi^+$ 
would be larger than that of
$\overline B\to \pi^-\pi^+$ which is consistent with a recent
observation~\cite{CLEOetapX}. 
Further, the penguin amplitude
in $b \to d$ processes may be enhanced such that direct
CP violation may become observable either inclusively or exclusively,
as in $B \to \pi \rho, \pi \omega,
\pi a_1, 3\pi, B^0 \to \pi^+ \pi^-$. Also, the recently observed
large value of $B(B^-\to\eta'K^-)$~\cite{CLEOetapX,CLEOetapK} 
and $B(B^-\to\eta' X; P_{\eta'}>2.2\;{\rm GeV/c})$~\cite{CLEOetapX}
may be relevant in this
context. Many CP studies with such rare decay modes and similar 
ones will have to be rethought.

Estimates of non-perturbative QCD effects are important to reliably 
compute
the $B_s - \overline B_s$ width difference~\cite{bbd} and the
inclusive, mixing-induced CP violating effects in
$B_d$ decays governed by $b \to u \overline u d, c \overline c d$
transitions~\cite{bbdinclusive}.
Superb vertex detectors would still be able to
isolate the inclusive $b \to u \overline u d$ transitions but
the signal of singly detached vertices may involve a larger
background than previously appreciated.

\vspace{1cm}
\centerline{\large\bf Acknowledgements} 
\vspace{0.5cm}

It is a pleasure to thank P. Ball, M. Beneke, G. Buchalla, 
F.E. Close, A. Falk, A. Kagan, 
J. Kuti, C. Morningstar, P. Page and H. Quinn
for discussions, and T. Browder and G. Buchalla for comments on an 
earlier draft. Many thanks go to Lois Deringer who diligently 
typed up this manuscript. This work was supported in part
by the Department of Energy, Contract No. DE-AC02-76CH03000 and
Grant No. DE-FG02-91ER40654.

\vspace{1cm}
\centerline{\large\bf Appendix}   \eqnapp%
\vspace{0.5cm}

The ALEPH measurement of
$\overline B\to D\overline DX$\cite{barate} is
\begin{equation}
B(\overline B\to D^0\overline D^0X, D^0 D^-X,
                 D^+\overline D^0X)
 = 0.128\pm   0.027 \pm 0.026\;.
\end{equation}
In order to obtain
$B(\overline B\to \overline DX)$, we need to add
$B(\overline B\to D^+D^-X)$ and
$B(\overline B\to D_s^+\overline D X)$
($\overline B\to N_c\; \overline N\;\overline D X_s$ is
  kinematically forbidden and
$\overline B\to N_c\;\overline N\;\overline D X$ is negligible).
The total $D^+D^-X$ production can be evaluated from ALEPH's
measurements~\cite{barate} by assuming
factorization and isopin symmetry\cite{recal} to be
$1\pm 0.4\%$,
where we have assigned a conservative error since
the assumption of factorization may not hold.

Our estimate for $D_s^+$ production in $b\to c\bar c s$ processes is
small.
In fact, the measured total $D_s^+$ production in tagged $\overline B$
decays is $Y_{D_s} f_{D_s} = 2\pm1\%$ (Tables I -- III), which
informs about the probability $P(b \to c\to D_s^+)$. Since about 10\%
of all $\overline B$'s decay as $b \to c + \overline D X_s$, and the 
formation
of $D_s^+$ from the $c$ quark entails phase-space suppression [due to
the existence of the two charmed mesons and two extra strange quarks
in the final state], we estimate that $B(\overline B\to D_s^+\overline D
X)$ not to exceed significantly the permille level.
Correcting for the key charm decay branching fractions
adopted in this note, we then obtain Eq. (\ref{eq:DbarALEPH}).

$B(\stackrel{(-)} {B} \to N_c X)$ can be related to the measurements on
$\Lambda_c$ using a model~\cite{recal,distw}. The assumptions of the model
are: (1) in charmed baryon production governed by the $b\to c
\overline q q'$ transition, the two quarks
$cq'$ end up in a single (excited) charmed baryon, (2) excited
$\Xi_c$ will end up as $\Xi_c$ and excited $\Lambda_c$ (or
$\Sigma_c$) will end up as $\Lambda_c$, and (3) the ratio of
$s\overline s$ pair creation to $u\overline u$ or $d\overline d$ pair
creation
is universal. The estimate for $B(\stackrel{(-)} {B} \to N_c X)$ is listed
in Table V. The predicted $\Xi_c$ production is found to be
much smaller than the measurement, and when any of the assumptions
are relaxed toward more realistic ones, the prediction becomes even
smaller.  Following the ideology presented in Refs.~\cite{recal,distw}, 
the results of the model can be interpreted therefore as 
model-independent upper limits on strange charm baryon yields in $B$ 
decays.

There exists another minor modification.
Refs.~\cite{recal,distw} claim that one must reassess the currently
accepted value
of $B(\Lambda_c \to pK^- \pi^+ )= 0.044 \pm 0.006$ because it has
been based on a
flawed model for $\overline B \to N_c X$. The model is
invalidated if
sizable $\overline B\to D^{(*)} \stackrel{(-)}{N}X$ are observed,
which were
predicted from simple Dalitz plot arguments.
Refs.~\cite{recal,distw} thus argue to use~\cite{shipsey}
\begin{equation}
 B(\Lambda_c \to pK^-\pi^+ )= 0.060 \pm 0.015 \;,
\end{equation}
a value adopted throughout this note.

\newpage

\vskip 1.5in

\begin{figure}
 \centering
 \mbox{\psfig{figure=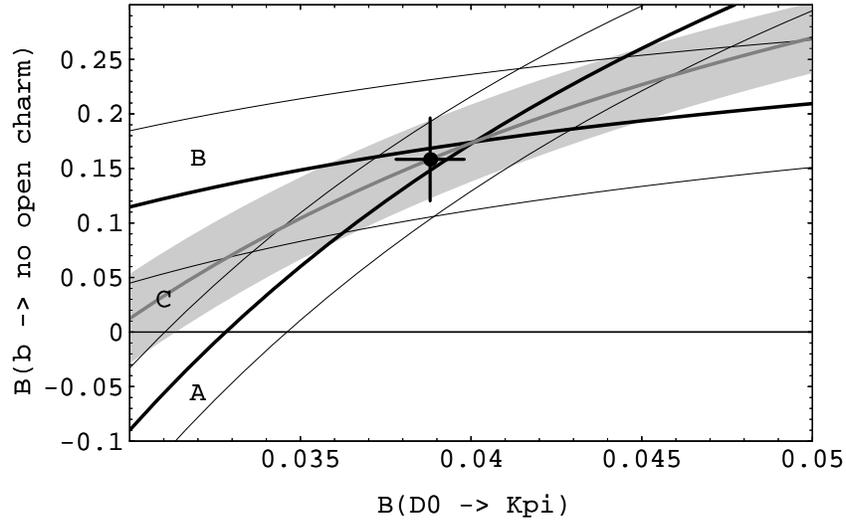,height=2.8in,width=4.4in}}
 \caption{Methods (labeled A, B, C; see text for detail)
   of estimating $B(b\to$ no open charm) are plotted against
   $B(D^0\to K^-\pi^+)$ together with bands corresponding to
   one standard deviation using CLEO data. 
   The point with error bars shows
   the world average of $B(D^0\to K^-\pi^+)$ and the best estimate of
   $B(b\to$ no open charm) via method C. }
 \label{fig:Bnoopnc}
\end{figure}

\begin{figure}
 \centering
 \mbox{\psfig{figure=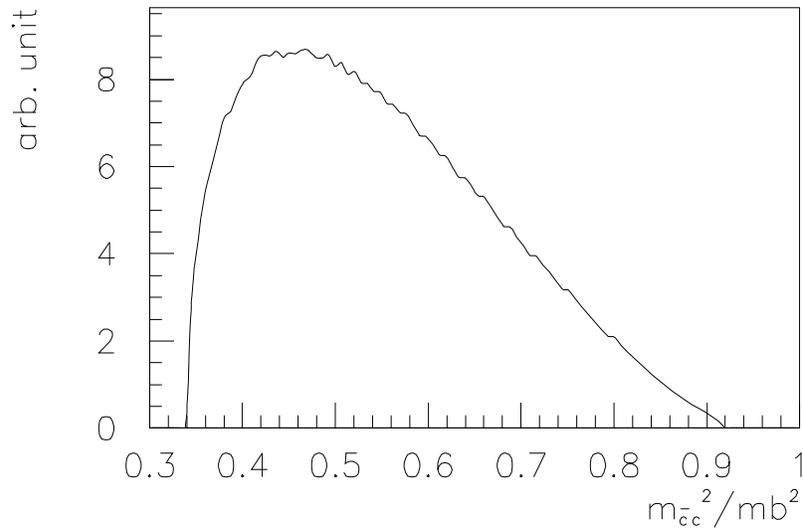,height=2.8in,width=4.2in}}
 \caption{The invariant mass distribution of the $c\bar c$ pair
       in the decay $b\to c\bar cs$~\protect\cite{bdy}.}
 \label{fig:mccbar}
\end{figure}

\begin{figure}
 \centering
 \mbox{\psfig{figure=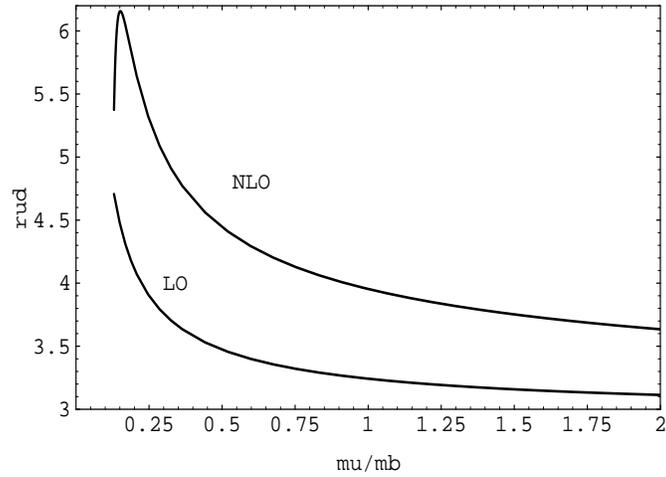,height=3.1in,width=4.5in}}
 \caption{Scale dependence of the $b\to c\bar ud'$ rate
  normalized to the semileptonic rate ($r_{ud}$) for the leading-order 
(LO) and 
  the next-to-leading-order (NLO) approximations~\protect\cite{rud}.}
 \label{fig:rudmu}
\end{figure}

\begin{figure}
 \centering
 \mbox{\psfig{figure=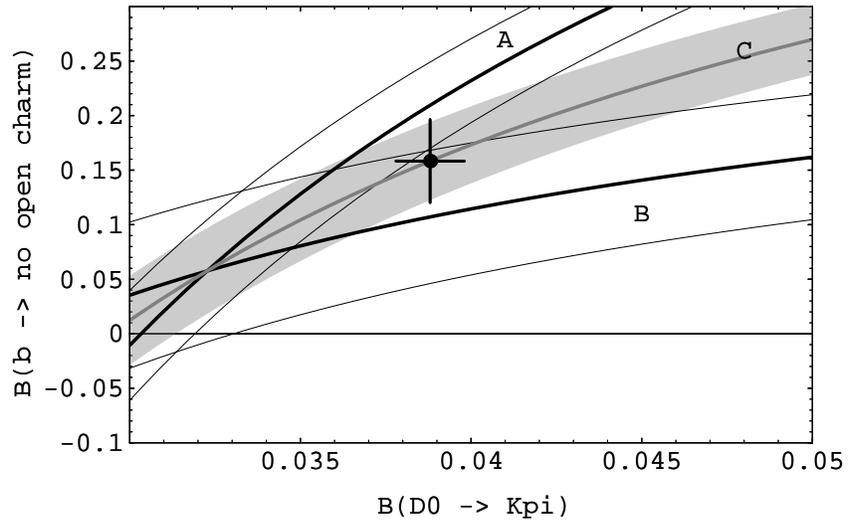,height=2.8in,width=4.4in}}
 \caption{Same as Figure 1, except for the value of $r_D$ which
 is hypothetically taken to be $0.20\pm0.03$.}
 \label{fig:Bnoopnc2}
\end{figure}

\begin{figure}
 \centering
 \mbox{\psfig{figure=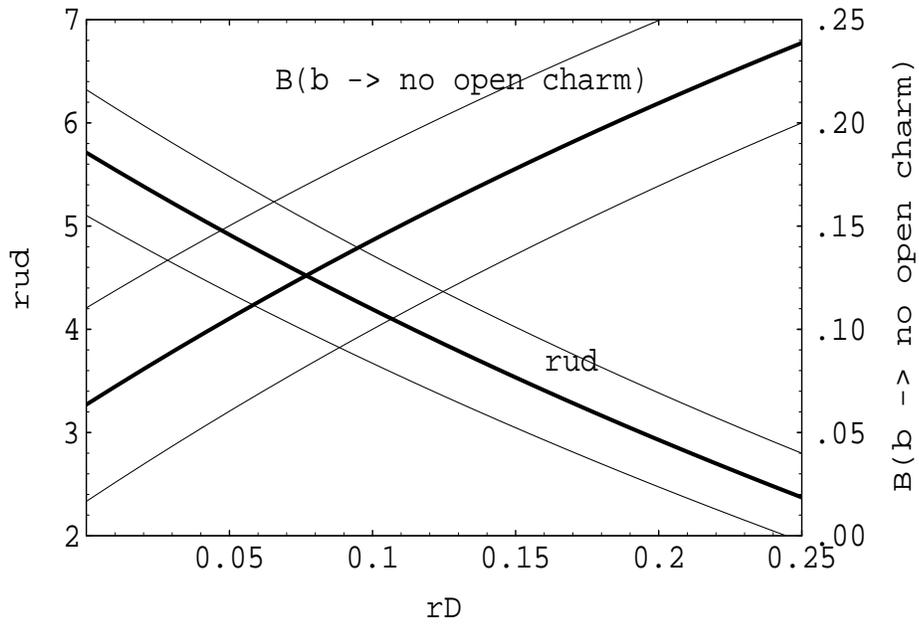,height=3.6in,width=5.0in}}
 \caption{$B(b\to\hbox{no open charm})$ (method A) 
     and $r_{ud}$ as functions of $r_D$.
     The inputs are essentially experimental only.}
 \label{fig:bnocrud}
\end{figure}

\newpage

\begin{table}
\caption{Inclusive Charmed Hadron Production in $B$ Meson
Decays as Measured by CLEO}
\begin{tabular}{|c|c|c|}
$T$ & $Y_T \equiv B(\overline B\rightarrow TX) +
B(\overline B\rightarrow \overline T X)$ & Reference \\
\tableline
$D$ & $(0.876 \pm 0.037) \left[\frac{0.0388}{B(D^0 \rightarrow
K^-\pi^+)}\right]$ & \cite{CLEODX} \\
\tableline
$D_s$ & $(0.1177 \pm 0.0093) \left[\frac{0.036}{B(D_s \rightarrow
\phi\pi )}\right]$ & \cite{ds} \\
\tableline
$\Lambda_c$ & $(0.030 \pm 0.005) \left[\frac{0.06}{B(\Lambda_c
\rightarrow pK^-
\pi^+)}\right]$ & \cite{baryonc}
\end{tabular}
\end{table}

\begin{table}
\caption{Absolute Branching Ratios of Key Charm Decays as Used in
this Note}
\begin{tabular}{|c|c|c|}
Quantity & Value & Comment \\
\tableline
\tableline
$B(D^0 \to K^-\pi^+)$ & 0.0388 $\pm$ 0.0010 & World Average
\cite{richman} \\
\tableline
$r_+ \equiv \frac{B(D^+\to K^- \pi^+ \pi^+)}{B(D^0 \to K^- \pi^+)}$ &
2.35 $\pm$
0.23 & CLEO \cite{rp} \\
\tableline
$r_s \equiv \frac{B(D_s \to\phi\pi )}{B(D^0 \to K^- \pi^+ )}$ & 0.92
$\pm$ 0.23 &
CLEO \cite{rs} \\
\tableline
$B(\Lambda_c \to pK^- \pi^+ )$ & 0.060 $\pm$ 0.015 & CLEO
\cite{shipsey}, see Appendix
\end{tabular}
\end{table}

\begin{table}
\caption{Inclusive Charmed Hadron Production in Tagged $B$ Decays as
Measured by
CLEO}
\begin{tabular}{|c|c|c|}
Observable & Value & Reference \\
\tableline
\tableline
$r_{\Lambda_c} \equiv \frac{B(\overline B\rightarrow
\overline\Lambda_c X)}{B(\overline
B\rightarrow \Lambda_c X)}$ & $ 0.20 \pm 0.14$ & \cite{baryonc}
\\
\tableline
$r_D \equiv \frac{B(\overline B\rightarrow \overline DX)}{B(\overline
B\rightarrow
DX)}$ & $0.107 \pm 0.034 $ & \cite{moriond} \\
\tableline
$f_{D_s} \equiv \frac{B(\overline B\rightarrow D^+_s X)}{Y_{D_s}}$ &
$0.172 \pm 0.083$ & \cite{ds}

\end{tabular}
\end{table}

\begin{table}
\caption{Charm Multiplicity in $B$ Meson Decays at $\Upsilon(4S),$
$Y_T \equiv
B(\overline B\to TX) + $ $B(\overline B\to \overline TX),$ and in
$b$-Hadron
Decays at $Z^0,
Y_T \equiv B(b\to TX)+B(b\to \overline TX)$}
\begin{tabular}{|c|c|c|c|}
Quantity & CLEO \cite{CLEODX} & ALEPH \cite{ccaleph} & OPAL
\cite{ccopal} \\
\tableline
\tableline
$(Y_D +Y_{D_s}) \frac{B(D^0 \to K^-\pi^+)}{0.0388}$ & 0.99 $\pm$ 0.04
& 1.01 $\pm$
0.05 & 0.93 $\pm$ 0.06 \\
\tableline
$Y_{\Lambda_c} \frac{B(\Lambda_c \to pK^- \pi^+ )}{0.06}$ & 0.030
$\pm$ 0.005 &
0.08 $\pm$ 0.01 & 0.09 $\pm$ 0.02 \\
\tableline
$Y_D +Y_{D_s}+Y_{\Lambda_c}$ & 1.02 $\pm$ 0.05 & 1.09 $\pm$ 0.07 &
1.02 $\pm$ 0.08
\end{tabular}
\end{table}

\begin{table}
\caption{Charmed Baryon [$N_c \equiv \Lambda_c, \Xi_c, \Omega_c$]
Production in $B$ Meson Decay as Predicted In
Refs.~\protect\cite{recal,distw}.}
\begin{tabular}{|c|c|}
Quantity & Value \\
\tableline
\tableline
$B(\overline B\to N_c X)$ & (0.0365 $\pm$ 0.0065)$
\;\left[\frac{0.06}{B(\Lambda_c \to pK^- \pi^+ )}\right]$ \\
\tableline
$B(\overline B\to\overline{N}_c X)$ & (0.0059 $\pm$ 0.0038)
$\;\left[\frac{0.06}{B(\Lambda_c \to pK^-\pi^+ )}\right]$ \\
\tableline
$Y_{N_c}$ & (0.0424 $\pm$ 0.0082)
$\;\left[\frac{0.06}{B(\Lambda_c \to
pK^- \pi^+ )}\right]$
\end{tabular}
\end{table}

\end{document}